\documentclass{article}

\usepackage{arxiv}

\usepackage[utf8]{inputenc} 
\usepackage[T1]{fontenc}    
\usepackage{hyperref}       
\usepackage{url}            
\usepackage{booktabs}  
\usepackage{amsmath}
\usepackage{array}      
\usepackage{amsfonts}       
\usepackage{nicefrac}       
\usepackage{microtype}      
\usepackage{lipsum}		
\usepackage{graphicx}
\usepackage{subcaption}
\usepackage{float}
\usepackage{multirow}

\usepackage{comment}
\graphicspath{{figures/}}
\usepackage{todonotes}
\usepackage{color}

\title{Is Time to Intervention in the COVID-19 Outbreak Really Important? A Global Sensitivity Analysis Approach}

\author{ \href{http://orcid.org/0000-0003-2103-6478}{\includegraphics[scale=0.06]{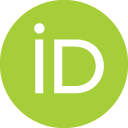}\hspace{1mm}Xuefei~Lu}\thanks{The authors contributed equally to this work.} \\
	Department of Decision Sciences\\
	Bocconi University\\
	Via Roentgen 1, Milan, Italy 20136 \\
	\texttt{xuefei.lu@unibocconi.it} \\
	\And
	\href{http://orcid.org/0000-0001-8659-6017}{\includegraphics[scale=0.06]{orcid.png}\hspace{1mm}Emanuele~Borgonovo} \\
	Department of Decision Sciences and BIDSA\\
	Bocconi University\\
	Via Roentgen 1, Milan, Italy, 20136 \\
	\texttt{emanuele.borgonovo@unibocconi.it} \\
}

\begin{document}
\maketitle

\begin{abstract}
	Italy has been one of the first countries timewise strongly impacted by the COVID-19 pandemic. The adoption of social distancing and heavy lockdown measures is posing a heavy burden on the population and the economy. The timing of the measures has crucial policy-making implications. Using publicly available data for the pandemic progression in Italy, we quantitatively assess the effect of the intervention time on the pandemic expansion, with a methodology that combines a generalized susceptible-exposed-infectious-recovered (SEIR) model together with statistical learning methods. The modeling shows that the lockdown has strongly deviated the pandemic trajectory in Italy. However, the difference between the forecasts and real data up to 20 April 2020 can be explained only by the existence of a time lag between the actual issuance date and the full effect of the measures. To understand the relative importance of intervention with respect to other factors, a thorough uncertainty quantification of the model predictions is performed. Global sensitivity indices show that the the time of intervention is 4 times more relevant than quarantine, and eight times more important than intrinsic features of the pandemic such as protection and infection rates. The relevance of their interactions is also quantified and studied.
\end{abstract}

\section{Introduction}
The COVID-19 epidemic initially developed in the Chinese region of Wuhan \cite{Wu2020265} before becoming a pandemic. As of public records on 21 April 2020, 2,478,634 people have been affected internationally, and more than 26\% of the world population is involved in some form of shelter-in-place or lockdown order. In the public political debate and in the scientific debate as well, great attention is posed on the effectiveness of intervention measures.
Time-wise, Italy has been one of the first Countries, after China and Iran, significantly hit by the pandemic and then forced to resort to the implementation of strictly increasing quarantine measures by an exponential increase in the COVID-19 infection. The situation in Italy up to the first half of March is well described in \cite{RamuRamu20,Grasselli2020}. Embracing the recommendations of the scientific experts (see also the considerations in \cite{RamuRamu20} concerning the exponential growth of the epidemic at the beginning of March 2020), since March 08 and March 09 2020 the local (Lombardy) and national (Italy) governments have imposed severe containment measures (lockdown, for simplicity, henceforth). The measures have, indeed, impacted the progression of the epidemic and avoided the exponential growth. On 20 April 2020 Italy counts a total of 181,228 confirmed individuals, with 24,114 deaths and a number of currently infected individuals equal to 108,237. 

\cite{Ande20Lancet} poses the question of how country-based mitigation measures may influence the course of the COVID-19 epidemic world wide. We have seen alternative attitudes internationally, with some countries intervening with strict lockdown measures within a short amount of time since the insurgence of the epidemics, and other adopting a more stage-phased approach. 

Questions that emerge are, then, whether the effect of the intervention can be quantitatively observed from the data and whether its effect is immediate or delayed. In fact, it is natural to expect a time-lag between the moment an intervention is decided and the time in which it is actually taking in place. A the same time, the intervention time is not be the sole factor explaining a change in the pandemic time progression. The quarantine, the infection rate, the protection rate, the average latent time are factors determining an epidemic evolution that are expected to play a role. We then wish to quantify the relative importance of the intervention time with respect to these other factors.

To answer these questions, we resort to mathematical modelling coupled with global sensitivity analysis. Forecasting the expansion of the COVID-19 epidemics is subject of intensive research investigation \cite{Wu2020689,EnseKupf20}.
We investigate the behavior of the pandemic combining predictions from a generalized SEIR model recently used in \cite{Kucharski2020,Wang2020,Peng2020} to describe the COVID-19 pandemic in China with machine learning approaches to increase interpretability of the findings \cite{Salt19}. We fit the generalized SEIR model to data of the pandemic in Italy covering the period 24 February to 20 April. We then run an uncertainty analysis with 1,000,000 Monte Carlo simulations and apply global sensitivity measures to quantify the relative importance of the factors mentioned above. We employ a replicated finite-difference decomposition method to study the intensity and sign of factor interactions.

\section{Results}
\paragraph{Time Delay of Intervention Effects.}
We start with results for the effects of the intervention time.
\begin{figure}[H]
	\centering
	\includegraphics[width = \textwidth]{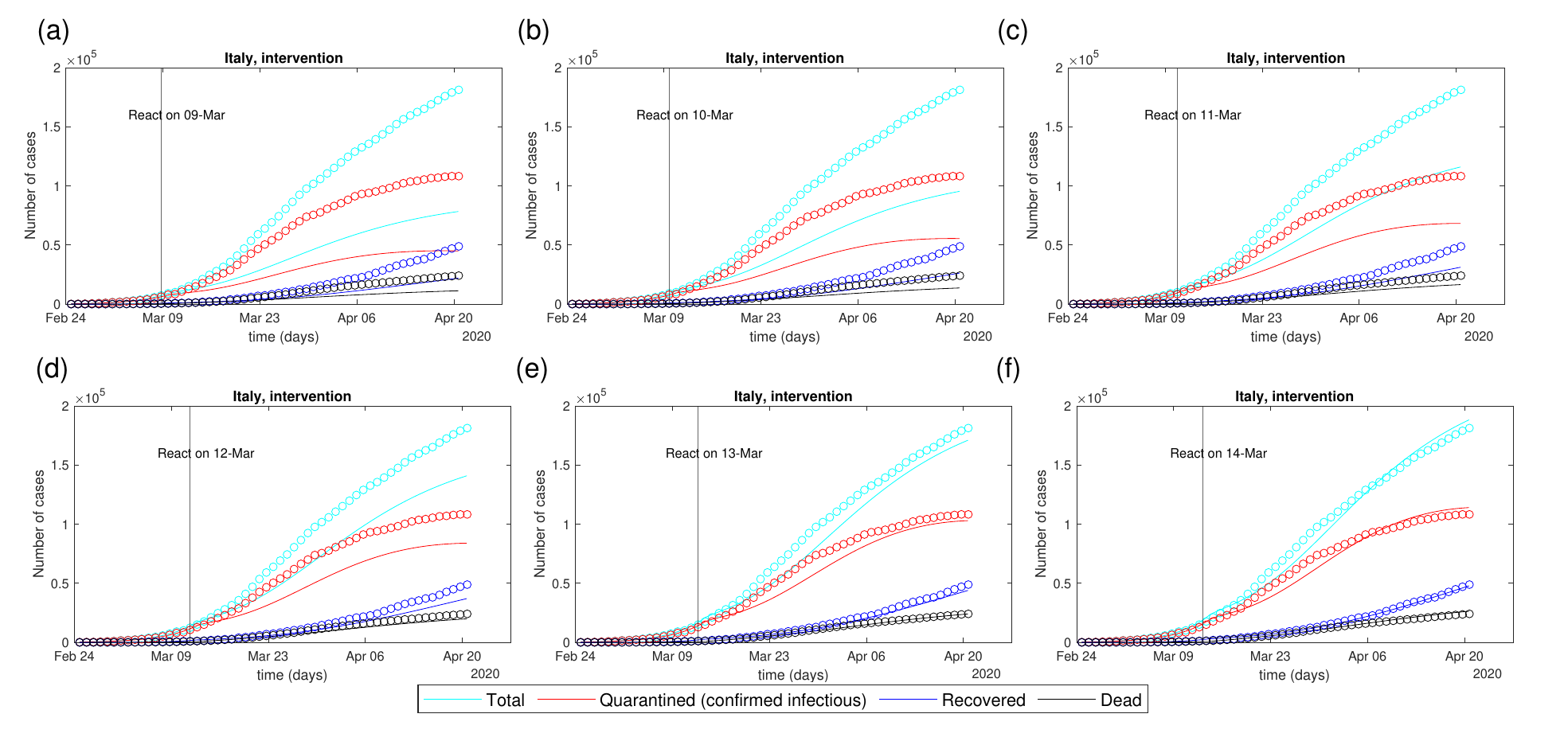}
	\caption{Significance of the intervention time, Italy}
	\label{fig:interventiondelay_Italy}
\end{figure}
Figure ~\ref{fig:interventiondelay_Italy} shows the following. The graphs represents the dynamic evolution of the pandemic comparing real data (circles) against the generalized SEIR model forecast (continuous line). The first graph report the results obtained for the case in which the intervention is modeled as fully active in the first day. We register a poor fit ($R^2=0.3861$), with the forecast largely underestimating the number of totally infected individuals (the root mean squared error (RMSE) is $RMSE=1.7416\times 10^4$). The second graph shows reports the comparison in the case the intervention is assumed to have effect 1 day later than the official issuance date. We register better fit, confirmed by an increase of $R^2$ from $0.3861$ to $0.5726$ (the RMSE is now $1.4611\times 10^4$, with a decrease of $16\%$). Then, graphs (c) to (f) show the effect of simulating the intervention with one additional day of delay per graph. Note that by simulating the intervention as starting 5 days later than the actual issuance date, the fitting performance improves notably. We register $R^2 = 0.9912$ and $RMSE=2.1365\times 10^3$, with a decrease of $87.73\%$ with respect to graph (a). 

Due to the fact that intervention time is not the only factor influencing the forecast, we also tried to tune the forecast by changing other parameters such as $I_0$ and $\delta$ and combinations, while leaving the intervention time at the issuance date. Our search did not lead to satisfactory results, signaling the intervention time as the main source of discrepancy. To confirm the relevance of the intervention time, we performed further analyses via global sensitivity measures.

\paragraph{Uncertainty quantification}
\begin{figure}[H]
	\centering
	\includegraphics[width = 0.5\textwidth]{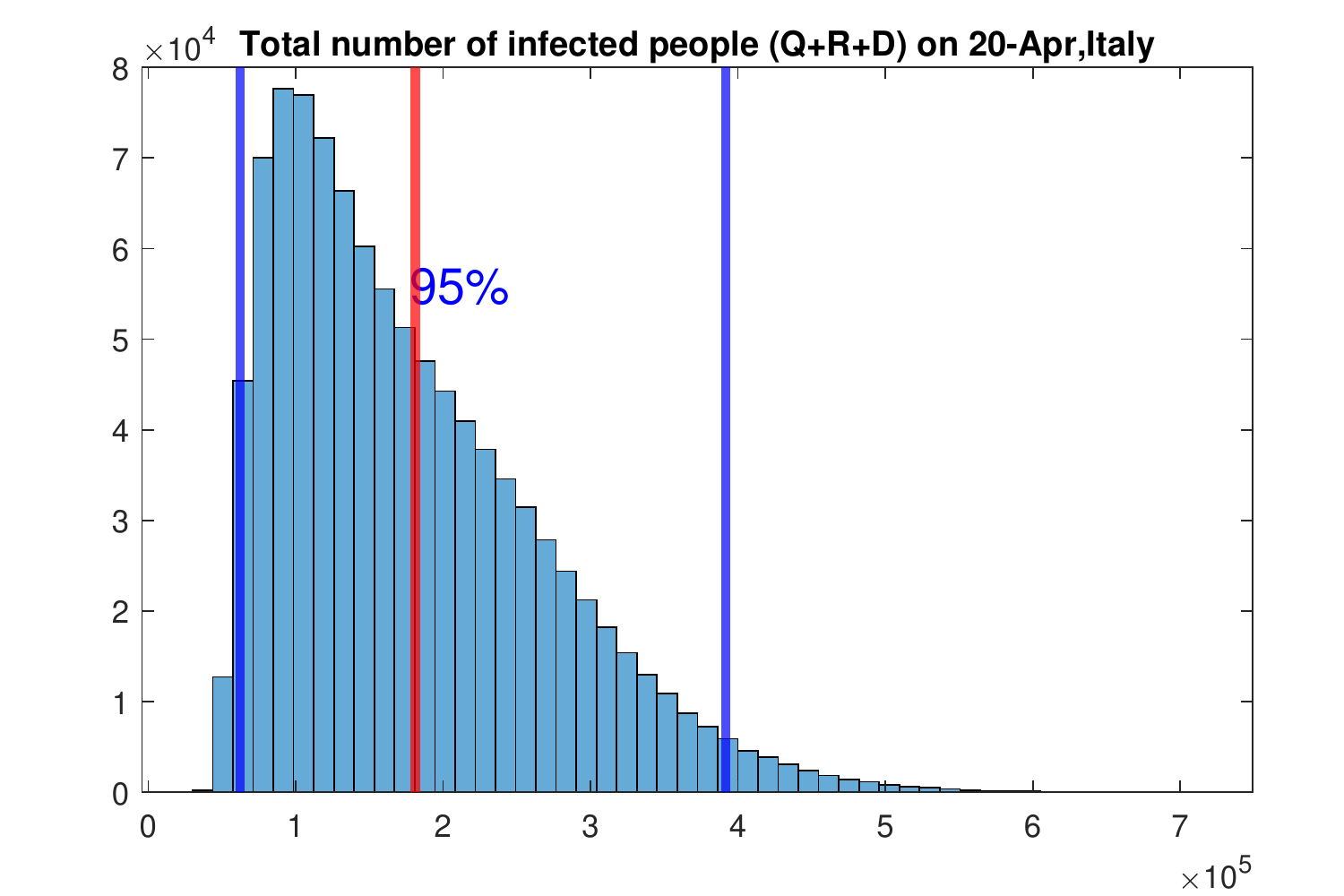}
	\caption{Empirical density  under uncertainty on 20 April in Italy; Blue lines are $2.5^{th}$ and $97.5^{th}$ percentiles; Red lines are actual confirmed cases. }
	\label{fig:hist_output}
\end{figure}
Figure \ref{fig:hist_output} displays the empirical density function for the SIER model forecast of the total number of infected individuals on April 20th, 56 days after the data collection start. The empirical density is skewed, with a mean of $1.7769\times 10^5$ individuals, standard deviation of $9.0035\times 10^4$, $95$ upper quantile on $3.5032\times 10^5$. Thus, the model foresees an expected number of infected individuals between $0.6216\times 10^5$ and  $3.9163\times 10^5$, which is consistent with actual data (red vertical line). 
The empirical density is computed on a sample of 1,000,000 realizations of the forecasts of total infected individuals. Uncertainty in the model predictions is the result of uncertainty in six factors acting as parameters of the generalized SEIR model: protection rate ($\alpha$), infection rate ($\beta$), average latent time ($\gamma^{-1}$), average quarantine rate ($\delta$), the number of initial infected people ($I_0$) and the intervention time (Interv.). The distributions are discussed in Section \ref{sec:method}. From these distributions uncertainty is propagated in the model through Monte Carlo simulation.

\paragraph{Factor Importance}
From the previous Monte Carlo sample, we have calculated global sensitivity measures that determine quantitatively the relative importance of the inputs [See Supplementary Appendix A]. Figure~\ref{fig:IM} shows the results.
\begin{figure}[H]
	\centering
	\includegraphics[width = 0.6\textwidth]{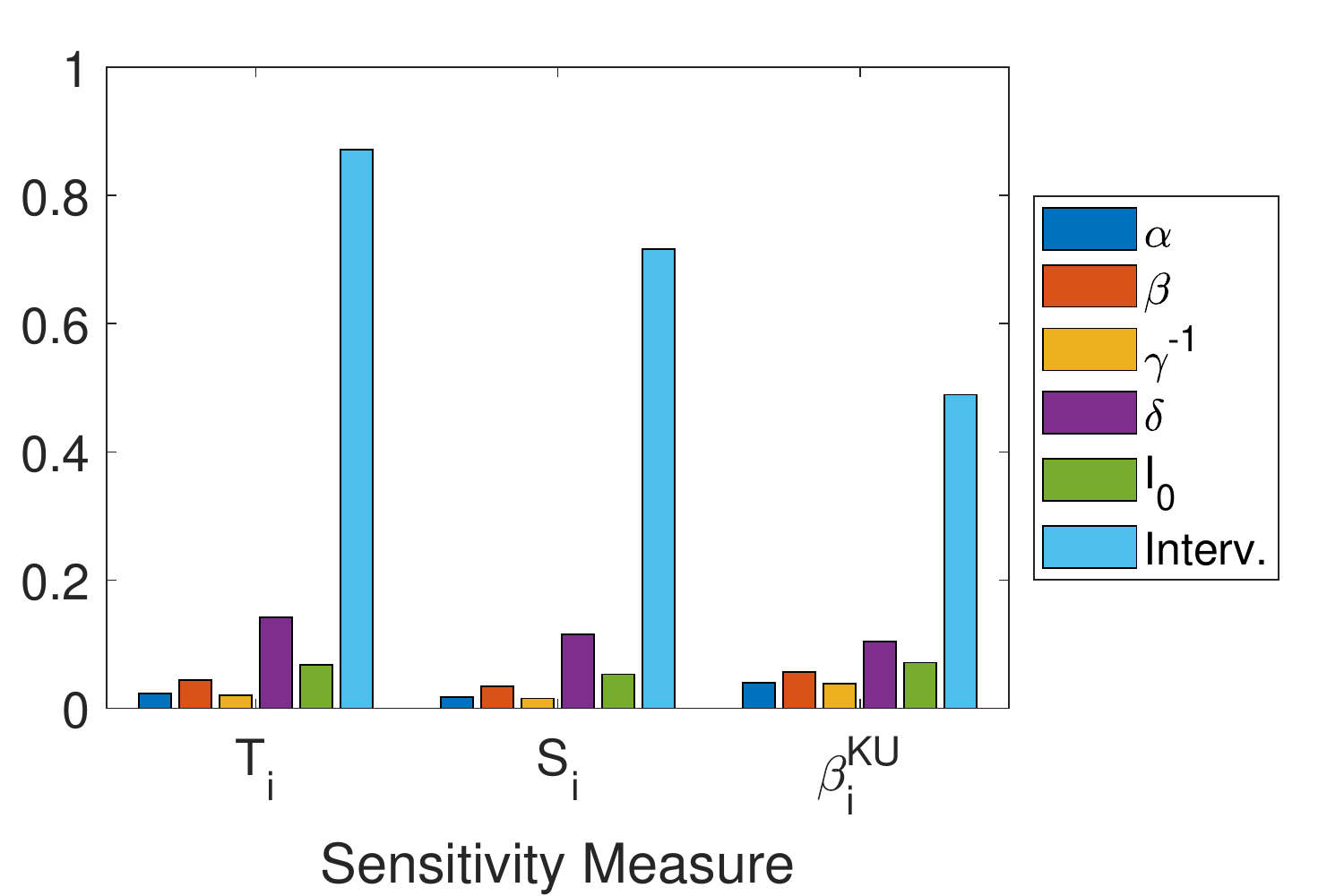}
	\caption{Importance of the six factors according to 3 sensitivity measures.}
	\label{fig:IM}
\end{figure}
Figure~\ref{fig:IM} shows that time to intervention (Interv.) is the key-driver in the variability of model predictions according to all the sensitivity measures used, i.e., variance-based total indices ($ T_i $), first order indices ($S_i$) and the distribution-based sensitivity measure ($\beta^{Ku}$) (see the supplementary quantitative appendix \ref{sec:Quant} for further details on the definition of these indices and the motivation for their simultaneous utilization.) It is followed by quarantine rate, number of initially infected individuals ($I_0$) and infection rate ($\beta$).  On a relative scale time to intervention is (depending on the sensitivity measure) $ 4.70 $ to $6.18$ times more important than the quarantine rate. In turn, the quarantine rate is $ 1.50 $ to $ 2.20 $ times more important than the initial number of affected individuals, which, in turn, is $ 1.30 $ to $1.60  $ times more important than the infection rate. Protection rate, infection rate and average latent time play an even minor role.   

\begin{figure}[H]
	\centering
	\includegraphics[width = 0.6\textwidth]{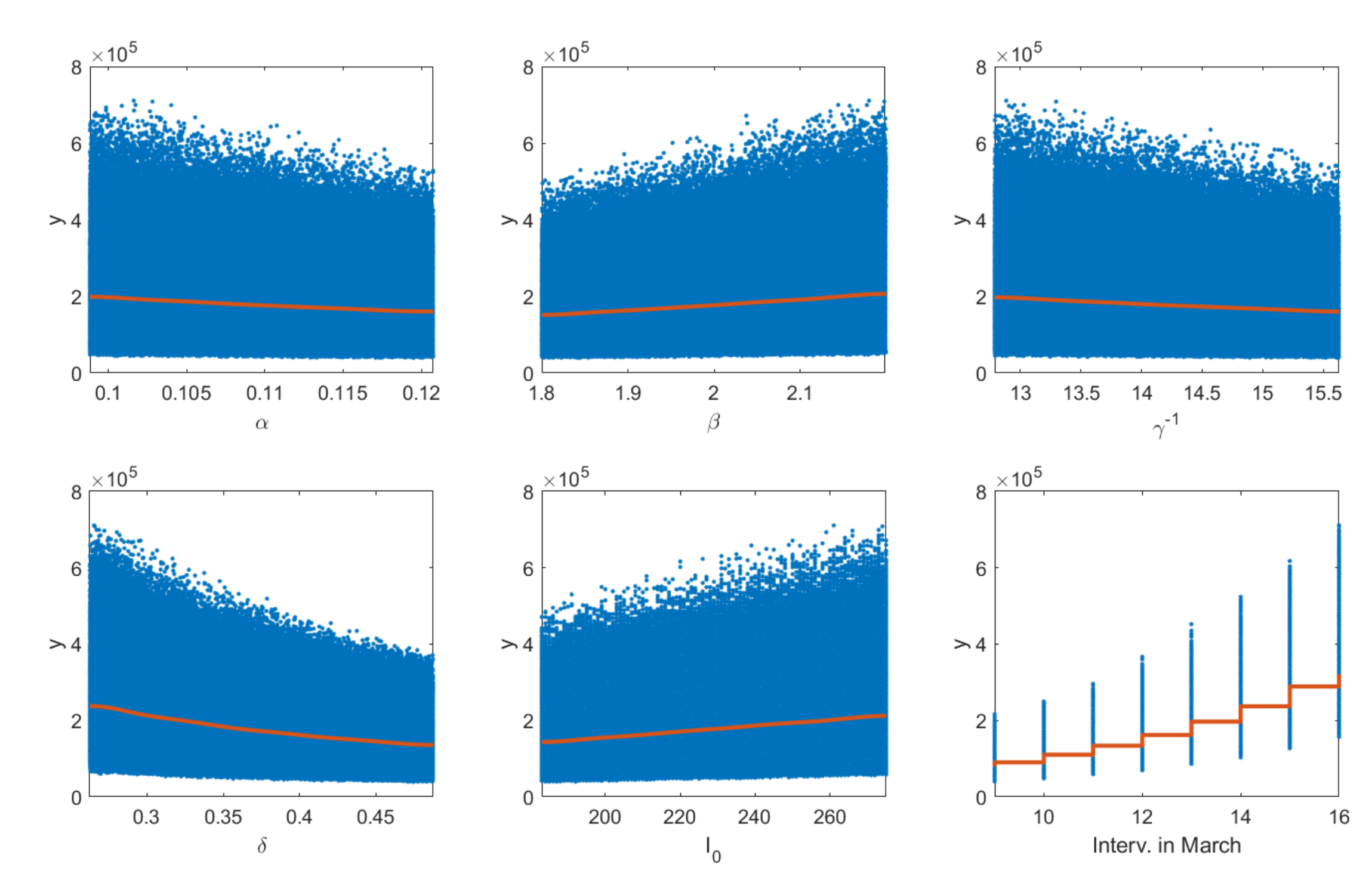}
	\caption{The conditional regression plots}
	\label{fig:cosi}
\end{figure}

\paragraph{Direction of influence}
Figure~\ref{fig:cosi} displays the conditional regression plots estimated from a subsample of size 500,000.
Figure~\ref{fig:cosi} shows the following. Each of the variables has a separately monotonic effect on the output. For instance, the expected total number of infected individuals is linearly and monotonically decreasing in the protection rate ($ \alpha $). Similarly, we expect a decrease associated with an increase in latent time and quarantine rate, but an increase associated with infection rate ($\beta$), in the number of initially infected individuals ($I_0$) and with the time of intervention (Interv.). These qualitative results are in accordance with intuition. The only difference is that the conditional regression function shows a (still decreasing) non-linear behaviour for the total number of infected individuals as a function of the quarantine rate. The last graph can be used to obtain a quantitative indication: when the intervention time is delayed by 7 days, the median total number of infected people would be expected to increase by around $300\%$. [This increase is calculated by comparing the median conditional on the intervention being effective on March 09 2020 to the median conditional the intervention being effective on March 16 2020.] 

\begin{figure}
	\centering
	\includegraphics[width = 0.6\textwidth]{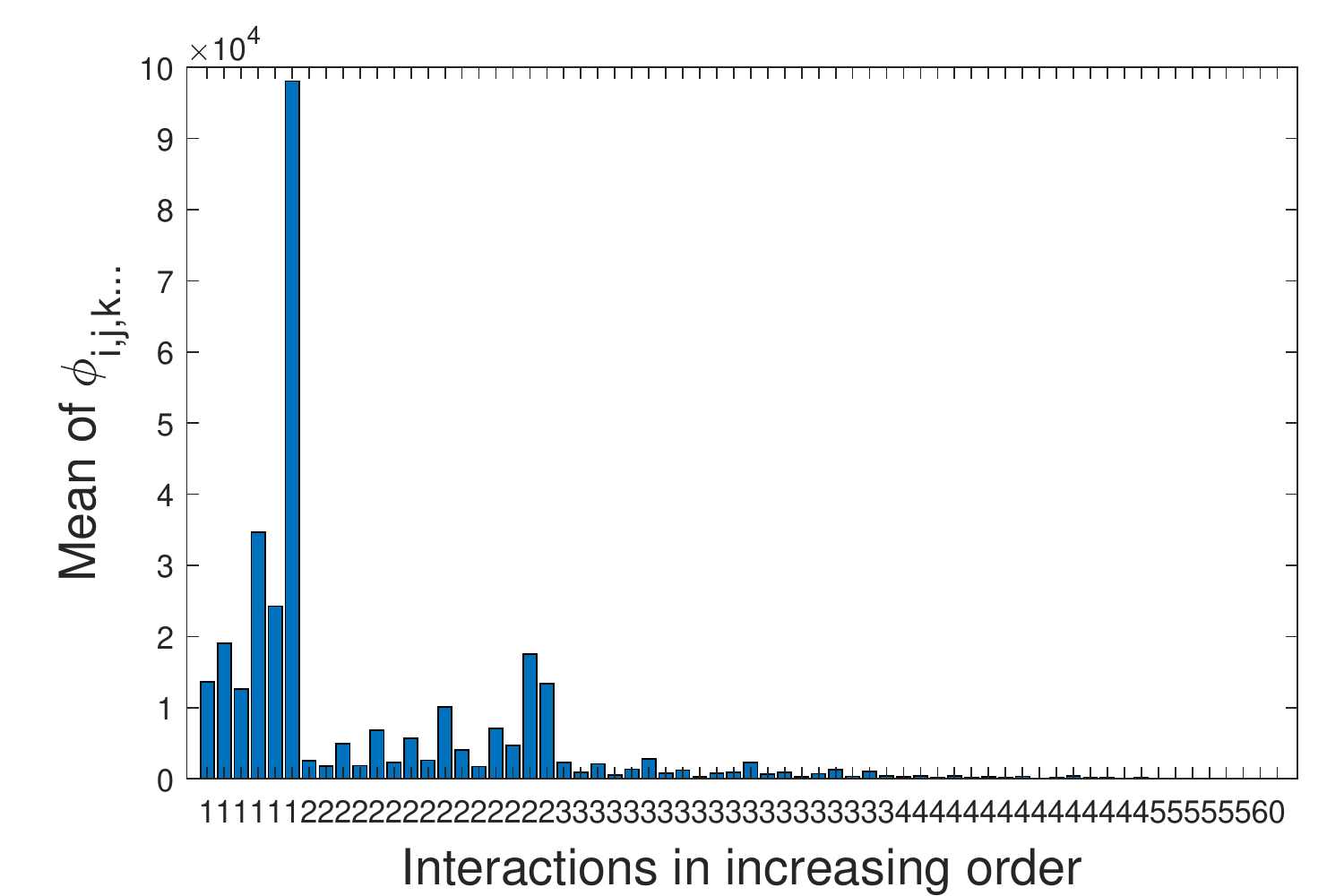}
	\caption{Averages of the magnitudes of interaction effects of all orders among the factors (The order ranges from 1 to 6). The first six bars concern the individual factor effects, the next 15 represent the average magnitudes of the second order effects, the next $20$ bars refer to third order effects, etc.. }
	\label{fig:interaction_mean}
\end{figure}

\paragraph{Interaction Quantification}
Figure~\ref{fig:interaction_mean} displays the magnitude of the interaction effects among the factors. The highest recorded interactions (the first six bars are individual effects) are among $\delta$, $I_0$ and the intervention time, namely the three most important factors. 
In terms of overall relevance, interaction effects account for less than $8\%$ of the variance of the affected individuals. The mean dimension is 1.1683 (see Appendix A for the mathematical definitions). This means interactions have a low overall effect, although the effect is not entirely negligible. In fact, the highest interaction (between intervention time and $\delta$) is $17.86\%$ of the highest individual effect.
Aside this interaction, we register a non-negligible interaction between intervention time and initial number of infected individuals ($I_0$). Note that these three factors are also the ones identified as most influential by the global sensitivity measures. 

\section{Discussion}
We have made inference on publicly available data of the COVID-19 pandemics in Italy through application of a SEIR model and the use of statistical techniques for sensitivity analysis.
The investigation has evidenced a time-lag between the moment interventions have been issued and the moment in which their effect has actually taken place changing the pandemic path. Indeed this result seem to reflect a physical mechanism. While shelter-in-place measures are officially established on a given date, it is unlikely that they have an immediate effect. For instance, individuals might not be able to interrupt all contacts immediately, they do not have instantaneous access to masks or other protecting equipment. Moreover, some activities cannot shutdown immediately due to technical reasons, and certain industrial compartments have to remain open to keep delivering essential services to the population. 

Our results have both policy-making and modeling implications. From a policy making viewpoint, global sensitivity measures indicate that the first two most important factors are the intervention time and the quarantine rate. These two factors relate to social distancing measures. Not only, but combined with the results of the time to intervention analysis, the findings indicate that the earlier intervention measures are introduced, but also the more rapidly (strongly) they can be implemented, the more effective they are in flattening the pandemic curve.  

From a modeling viewpoint, the strong impact of the intervention time suggests indeed that this variable needs to be taken into account to increase the prediction accuracy of SEIR models.

Regarding limitations of the present analysis, further work can be done to determine the distributions of the factors more precisely and, potentially, to include statistical correlations among the factors. Also, the analysis has been focused on the total number of infected individuals at 56 days from the initial observations. However, the analysis can also be made time dependent, varying the final time (e.g., using an analysis at 50 days or at 100 days). Of course, the longer the time horizon, the larger the uncertainty in the forecast we expect. Also, we have chosen one specification of the SEIR models and alternative specifications may be used.

The paper moves in the direction advocated by \cite{Salt19}. The uncertainty quantification is directly applicable to other SIER models and/or even to other type of models used to forecast or study the COVID-19 pandemics. Also, while we have analyzed data from Italy, the analysis can be replicated with data coming from other Countries and regions. 
The approach yields coherent insights not only regarding the most important inputs but also about the role of interactions and direction of change. This enriches the spectrum of insights gained from the model and the data with respect to current practice, with analyses that are limited only to varying one factor at a time. The analysis helps modelers in understanding key-factors of uncertainty increasing transparency. Collecting information on these factors might reduce uncertainty in the forecasts improving modelling efficacy and helping result communications to the public (see the recent critique in \cite{Bui2020}).

\section{Methods} \label{sec:method}
We consider a representative of the SEIR family of models. These models are a cornerstone in epidemiological studies \cite{Kermack1927,Hethcote2000} and have been used in the very recent studies of \cite{Kucharski2020,Wang2020} regarding the COVID-19 pandemics. 
We consider the version used to simulate the COVID-19 outbreak in China in the work of \cite{Peng2020}. This generalization allows a new quarantined state that includes the effect of soft measures before lockdown. In total, the following seven different states are considered: susceptible $S(t)$,  insusceptible $P(t)$, exposed (but not yet infectious) $E(t)$, infectious $I(t)$, quarantined $Q(t)$, recovered $R(t)$, deceased $D(t)$.

\begin{minipage}[b]{0.5\linewidth}	
	\includegraphics[width=\textwidth]{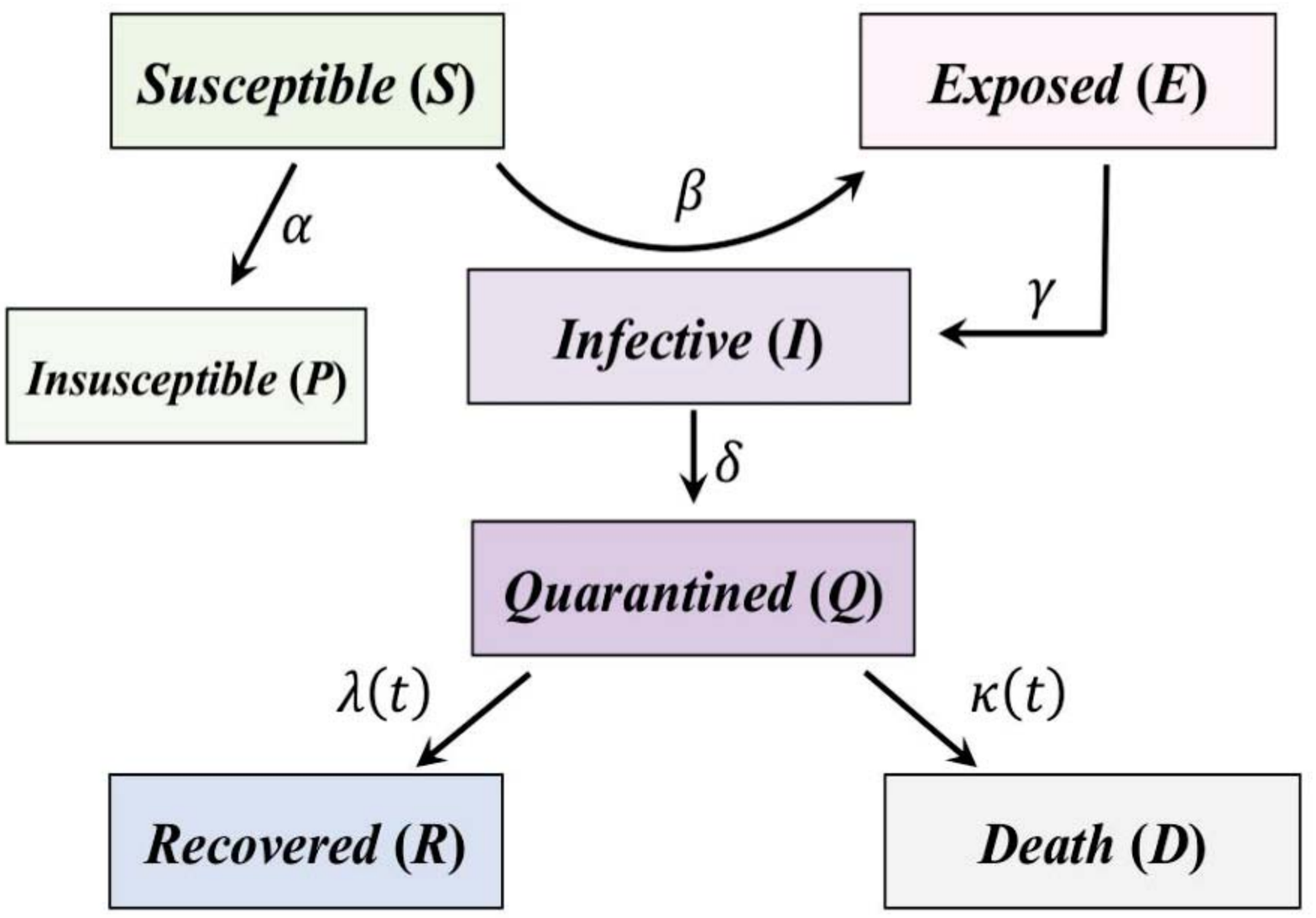}
	The generalised SEIR model. Adapted from \cite{Peng2020}.
\end{minipage}
\hfill
\begin{minipage}[b]{0.5\linewidth}
	\begin{eqnarray*}
		\frac{\mathrm{d}S(t)}{\mathrm{d}t} &=& -\alpha S(t) -\beta \frac{S(t)I(t)}{N_{pop}}  \\
		\frac{\mathrm{d}E(t)}{\mathrm{d}t} &=& -\gamma E(t) +\beta \frac{S(t)I(t)}{N_{pop}}  \\
		\frac{\mathrm{d}I(t)}{\mathrm{d}t} &=& \gamma E(t) -\delta I(t)  \\
		\frac{\mathrm{d}Q(t)}{\mathrm{d}t} &=& \delta I(t) -\lambda(t)Q(t) - \kappa(t) Q(t)  \\
		\frac{\mathrm{d}R(t)}{\mathrm{d}t} &=& \lambda(t) Q(t) \\
		\frac{\mathrm{d}D(t)}{\mathrm{d}t} &=& \kappa(t) Q(t) \\
		\frac{\mathrm{d}P(t)}{\mathrm{d}t} &=& \alpha S(t) 
	\end{eqnarray*}
\end{minipage}
The model includes the following parameters: i) protection rate: $\alpha$; ii)  infection rate, 
$\beta$; iii) average latent time: $\gamma^{-1}$;   iv) average quarantine rate: $\delta$; v) time-dependent cure rate coefficients: $\lambda_0$ and $\lambda_1$ (the time dependent cure rate is written as by $\lambda \left(t\right)=\lambda_0 \left(1-\exp \left(-\lambda_1 \;t\right)\right)$); vi) time-dependent mortality rate coefficients: $\kappa_0$ and $\kappa_1$ (the mortality rate curve is written as $\kappa \left(t\right)=\kappa_0 \exp \left(-\kappa_{1\;} t\right)$).
For simplicity,  the birth and natural death processes are not considered so that a constant population is assumed. The above model has been implemented in \textsc{Matlab}  \cite{Cheynet2020} and can be openly accessed at \href{https://github.com/ECheynet/SEIR}{https://github.com/ECheynet/SEIR}. 

\paragraph{Data Collection and Model Fitting}
We collected data from Italy's the Department of Civil Protection publicly available repository \href{https://github.com/pcm-dpc/COVID-19}{https://github.com/pcm-dpc/COVID-19}. 

\begin{figure}
	\centering
	\includegraphics[width = 0.5\textwidth]{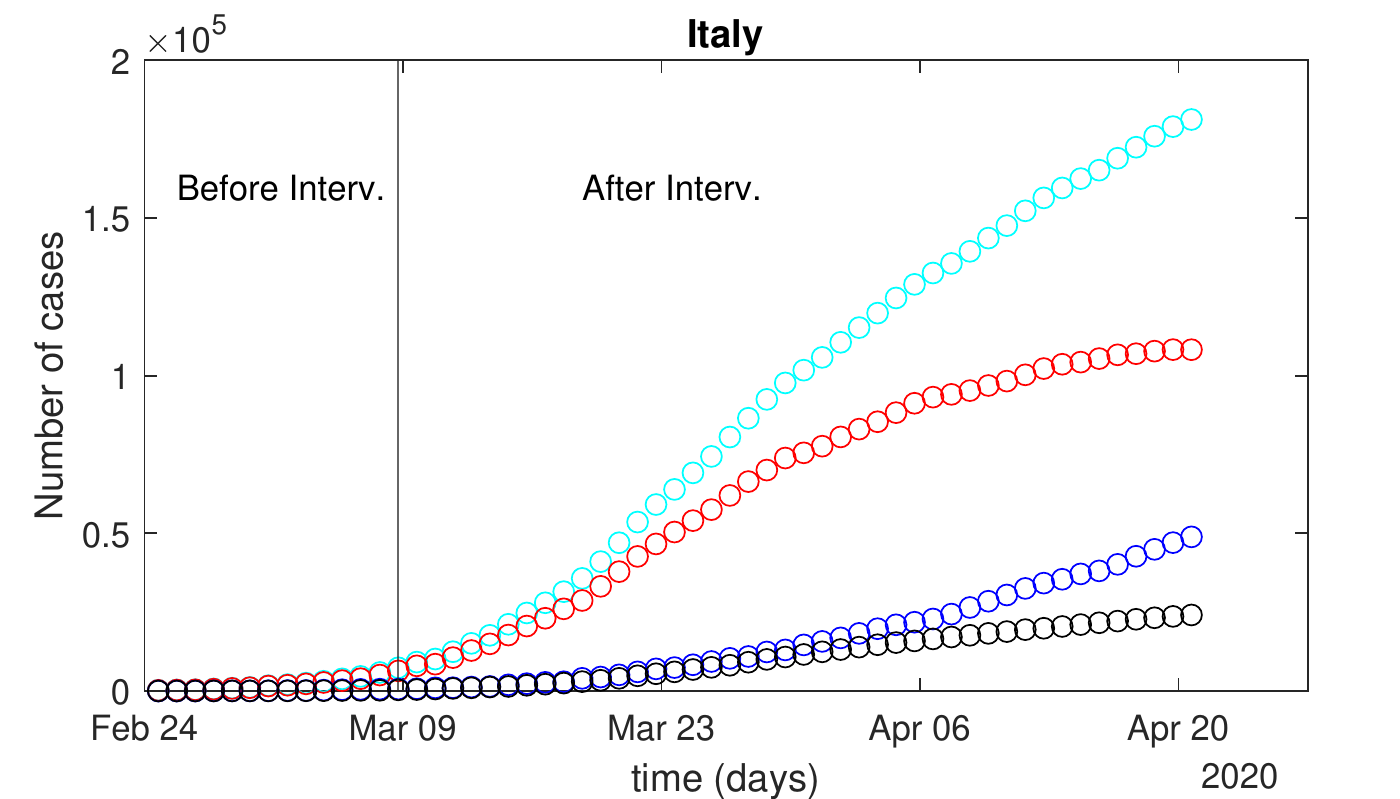}\\
	\includegraphics[scale=0.5]{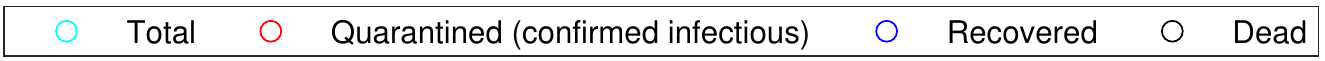}
	\caption{Data used for modelling, from 24 February 2020 to 20 April 2020.}
	\label{fig:data}
\end{figure}

Figure~\ref{fig:data} displays the number of total, quarantined, recovered, deceased individuals as a function of time, as per the publicly available data from 24 February till 20 April. The vertical lines display the dates of 09 March 2020, at which the lockdown measures were officially deployed. Fitting the generalized SEIR model discussed above on the data from 24 February to 08 March (13 days) for Italy, we obtain the pre-lockdown parameter estimates. Then, fitting the same model on the data from 09 March to 20 April (42 days), we obtain the after-lockdown parameter estimates reported in Table \ref{Tab:estimatedparameter}.  
\begin{table}[H]
	\centering
	\caption{Estimated Model Parameters for Italy \label{Tab:estimatedparameter}}
	\begin{tabular}{|l|l|l|l|l|l|l|}
		\hline
		& $\alpha$ & $\beta$ & $\gamma^{-1}$ & $\delta$ & $\lambda$ & $\kappa$ \\ \hline
		Pre-lockdown & 0.0000 & 1.1801 & 2.182 & 0.5985 & $[0.0437,0.1161]$ & $[0.0162,0.0461]$ \\ \hline
		After-lockdown & 0.1098 & 2.0000 & 14.2091 & 0.3750 & $[0.0167,2.0000]$ & $[0.0240,0.0432]$ \\ \hline
	\end{tabular}
\end{table}
Note that the after-parameters estimates are significantly different from the pre-lockdown estimates. For instance, the protection rate $\alpha$ increases by over 2,000 times; The average latent time increases by over $6.5$ times; the quarantine rate $\delta$ decreases by $37\%$. 


\begin{figure}[H]
	\centering
	\includegraphics[width =0.32\textwidth]{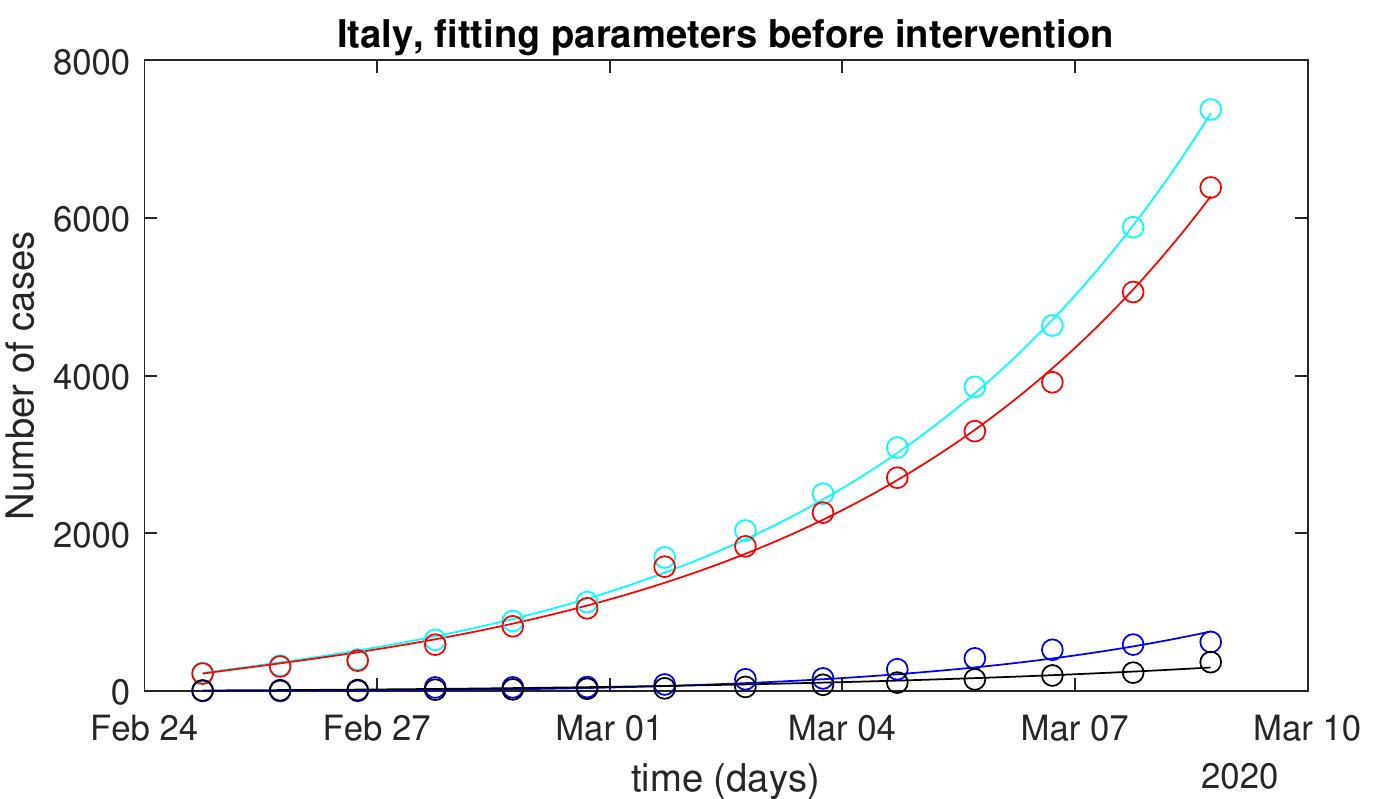}
	\includegraphics[width = 0.32\textwidth]{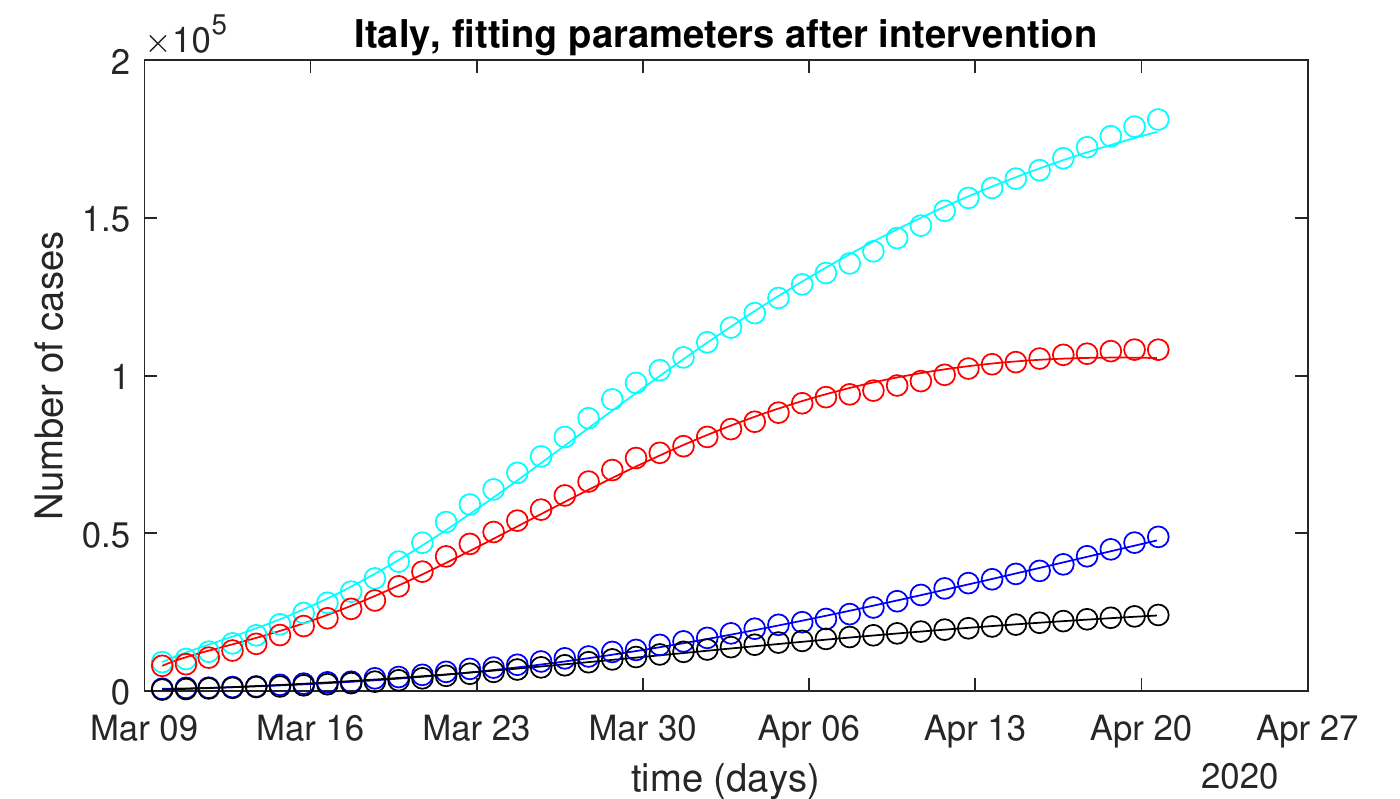}\\
	\includegraphics[scale=0.5]{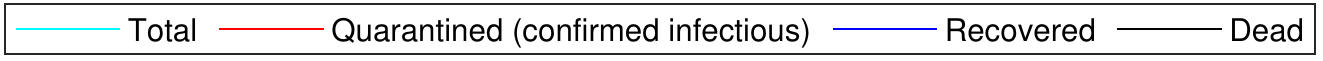}
	\caption{Model fitting before/after intervention, points are real data, lines are fitted.}
	\label{fig:data_fitting}
\end{figure}
Figure~\ref{fig:data_fitting} displays the total number of infected individuals calculated by the fitted models (continuous lines) against the real observations (circles). In terms of fitting performance, we register an average coefficient of determination ($R^2$) over the quarantined, recovered and deceased cases equals to $0.9542$ in the pre-intervention period, and $R^2 = 0.9980$ in the after-intervention period.

\paragraph{Uncertainty quantification}
To further understand the results, we have let the model undergo an uncertainty quantification in order to be able to compute global sensitivity measures. We used the parameter distribution assignments displayed in Table~\ref{Tab:inputdistribution} (see also \cite{Peng2020}) and generated a random Monte Carlo sample of size $ N=1,000,000 $. 
\begin{table}[H]
	\caption{Uncertainty quantification: the after lockdown parametric values of $\alpha$, $\beta$, $(\gamma)^{-1}$ and $\delta$ are assigned distributions as per the data of \cite{Peng2020}. The intervention time is assumed to be a uniformly distributed random variable in a one-week range after the actual issuance date. Independence is assumed. \label{Tab:inputdistribution}}
	\begin{tabular}{|c|c|c|c|c|}
		\hline
		& \multicolumn{1}{c|}{$\alpha$} & \multicolumn{1}{c|}{$\beta$} & \multicolumn{1}{c|}{$\gamma^{-1}$} & $\delta$ \\ \hline
		Distribution & \multicolumn{1}{c|}{Uniform} & \multicolumn{1}{c|}{Uniform} & \multicolumn{1}{c|}{Uniform} & Uniform \\ \hline
		Support & \multicolumn{1}{c|}{$ \alpha^{after}\pm 10\%$} & \multicolumn{1}{c|}{$ \beta^{after}\pm 10\%$} & \multicolumn{1}{c|}{$\gamma^{-1,after}\pm 10\%$} & $\delta^{after}\pm 30\%$ \\ \hline
		& $I_0$ & The intervention day & \multicolumn{2}{l|}{\multirow{3}{*}{}} \\ \cline{1-3}
		Distribution & Discrete uniform & Discrete uniform & \multicolumn{2}{l|}{} \\ \cline{1-3}
		Support & $I_0 \pm 20\%$ & $\{$09-Mar$ + z\},$ with $z=0,1,...,7$ & \multicolumn{2}{l|}{} \\ \hline
	\end{tabular}
\end{table}

From this sample, we have quantified the uncertainty about the forecasted number of infected individuals in Italy from 24 February to 20 April (56 days). 
Also, we have calculated importance indices that go under the name of global sensitivity measures to determine the quantitatively the relative importance of the inputs [See Supplementary Appendix A]. Specifically, we have used the first and total order global sensitivity measures based on variance contribution and a sensitivity measures based on the distance between cumulative distribution functions. The rationale of the choice is as follows. Regarding importance, the total order as well as the distribution-based sensitivity measure possess the properties that they are null if and only if the model output is probabilistically independent of the input of interest. Thus, their null value reassures us that the input is, indeed, not influential. Regarding interaction quantification, the difference between the first and the total order indices provides information about the relevance of interactions, giving insights on whether a factor's influence is due to its individual action or due to interactions with other factors. This insight, however, is an overall insight and does not reveal which interactions (if any) are important.
Towards this understanding we use a sample of size 20,000, replicating a full factorial design and decomposing the output changes via a finite difference operation (see Appendix \ref{sec:Quant} fpor further details). The scheme allows us to compute a total of 20,000 first order effects for each input, 20,000 second order interaction indices for all pairs, and 20,000 interaction indices of order 3, 4, and 5 and 6, enabling interactions to be thoroughly studied. 

From the original sample, we also computed conditional regression curves. These curves express the expected number of infected individuals as a function of each of the uncertain parameters. The graphs lead information about whether the increase in a parameter leads to an increase (decrease) of the number of infected individuals.

\appendix
\section{Supplementary Appendix 1: Quantitative Details on Sensitivity Analysis\label{sec:Quant}}

This section is divided into two parts. In the first part, we address the
definition of the global sensitivity measures used in this work. In the
second part we ketch the principles of their
estimation and list the corresponding subroutines.

\subsection{Sensitivity Measures: a Concise Overview}

For broad overviews on sensitivity analysis, we refer the reader
to \cite{Saltelli2008,Borgonovo2016book}. We concisely overview the methods used here. Let $y=g(\mathbf{x})$, where $g:\mathcal{X}\rightarrow \mathbb{R} $ is
a multivariate input-output mapping,  $\mathbf{x}=(x_1,x_2,...,x_n)$ be the
vector of the model inputs, with $ \mathbf{x}\in \mathcal{X} $ and $y$ the output of interest. 
Let also $ (\mathcal{X},\mathcal{B}(\mathcal{X}),\mathbb{P}) $ be the measure space of the model inputs and let $\mathbf{x}^{0}$%
, $\mathbf{x}^{1}$,..., $\mathbf{x}^{k}$,$\mathbf{x}^{k+1}$,..., $\mathbf{x}%
^{N}$ be a collection of points in $\mathcal{X} $ sampled from the input probability measure. Let $y^{0}=g(\mathbf{x}^{0})$, $y^{1}=g(\mathbf{x}^{1})$,...,$y^{N}=g(%
\mathbf{x}^{N})$ be the corresponding values of the model the output. Then, let $%
\Delta y^{k}=y^{k+1}-y^{k}$ denote the variation in the model output registered with the inputs shift from $\mathbf{x}^{k}$ to $\mathbf{x}^{k+1}$. We then have a sequence
of finite changes $\Delta y^{k}$, $%
k=0,1,2,...,N-1$. We can decompose each change in $ 2^n $ effects, writing: 
\begin{equation}
\Delta y^{k}=\sum_{i=1}^{d}\phi _{i}^{k\rightarrow k+1}+\sum_{i<j}\phi
_{i,j}^{k\rightarrow k+1}+\cdots +\phi _{1,2,...,d}^{k\rightarrow k+1},
\label{eq:finitchangedec}
\end{equation}%
where 
\begin{equation}
\begin{cases}
\begin{array}{c}
\phi _{i}^{k\rightarrow k+1}=g\left( x_{i}^{k+1}:\mathbf{x}_{-i}^{k}\right)
-g\left( \mathbf{x}^{k}\right) \\ 
\phi _{i,j}^{k\rightarrow k+1}=g\left( x_{i,j}^{k+1}:\mathbf{x}%
_{-\{i,j\}}^{k}\right) -g\left( \mathbf{x}^{k}\right) -\phi
_{i}^{k\rightarrow k+1}-\phi _{j}^{k\rightarrow k+1} \\ 
\dots.%
\end{array}%
\end{cases}
\label{eq:phiijk}
\end{equation}%
In eq. \eqref{eq:phiijk}, the point $\left( x_{i}^{k+1}:\mathbf{x}%
_{-i}^{k}\right) $ is obtained shifting $x_{i}$ from $x_{i}^{k}$ to $%
x_{i}^{k+1}$, while keeping the remaining inputs at their values in $\mathbf{%
	x}^{k}$. Similarly, $\left( x_{i,j}^{k+1}:\mathbf{x}_{-i,j}^{k}\right) $
denotes the point obtained by simultaneously shifting $x_{i}$ and $x_{j}$
from $x_{i}^{k},x_{j}^{k}$\ to $x_{i}^{k+1},x_{j}^{k+1}$ while keeping the
remaining inputs at their values in $\mathbf{x}^{k}$.

The terms $\phi _{i}^{k\rightarrow k+1}$ in equation (\ref{eq:finitchangedec}%
) are individual contributions, $%
\phi _{i,j}^{k\rightarrow k+1}$ second order contributions, quantifying the residual interaction between $%
x_{i}$ and $x_{j}$; a similar interpretation is shared by higher order terms. Dividing these second
order effects by $(x_{i}^{k+1}-x_{i}^{k})(x_{j}^{k+1}-x_{j}^{k})$, we obtain
the second order Newton ratios that provide indications about whether
interactions are locally positive (synergistic) or negative (antagonistic).

Sampling the model $ N $-times, we
obtain $N-1$ replicates of the finite change indices, $\phi_{i,j,...,m}^{k\rightarrow k+1}$. These sensitivity measures can be used, on
the one hand, to obtain detailed regional information on interactions. 
Assume the inputs are independent. Let $V_{Y}$ the variance of $Y$
induced by uncertainty in the model inputs. As proven in \cite{EfroStei81},
we can write 
\begin{equation}
V_{Y}=\sum_{i=1}^{d}V_{i}+\sum_{i<j}V_{i,j}+\cdots +V_{1,2,...,d}
\label{eq:sigma2Y}
\end{equation}%
where 
\begin{equation}
\begin{array}{c}
V_{i}=\int g_{i}(x_{i})^{2}dF_{i}(x_{i}) \\ 
V_{i,j}=\int g_{i,j}(x_{i,j})^{2}dF_{i,j}(x_{i},x_{j}) \\ 
...%
\end{array}%
\end{equation}%
and the functions $g_{i}(x_{i}),g_{i,j}(x_{i},x_{j})$ obtained by taking
appropriate conditional expectations \cite{EfroStei81,Borgonovo2016book}.  In equation (%
\ref{eq:sigma2Y}), $V_{i}^{2}$ is the portion of the variance attributed to $%
X_{i}$ alone, $V_{i,j}^{2}$ the portion of the variance attributed to the
residual interaction between $X_{i}$ and $X_{j}$. In the literature, one
considers the normalized version 
\begin{equation}
S_{i}=\dfrac{V_{i}}{V_{Y}},  \label{eq:Si}
\end{equation}%
called first order Sobol' indices. One also defines the total sensitivity
index of $x_{i}$, here denoted by $T_{i}$ \cite{Homma1996}. The index is
written as: 
\begin{equation}
T_{i}=\frac{V_{i}+\sum_{j=1,j\neq i}^{d}V_{i,j}+\cdots +V_{1,2,...,d}}{V_{Y}}%
.  \label{eq:tauisquare}
\end{equation}%
and represents the total portion of the variance of $Y$ associated with $%
x_{i}$. We note that by a result due to \cite{Owen03}, the sum of the total
order indices equals the mean dimension of $g$ in the superimposition sense.

There is a bridge between finite change sensitivity indices $\phi
_{i}^{k\rightarrow k+1}$ and the total index $T_{i}^{2}.$ In fact, let $\Phi
_{i}$ denote the random variable whose realizations are the $\phi
_{i}^{k\rightarrow k+1}$. Then, from the theory of the Sobol' estimation
design \cite{Sobo93} \cite{GambJano16} \cite{BorgRabi20}, we get 
\begin{equation}
T_{i}=\frac{\mathbb{V}\left[ \Phi _{i}\right] }{2}.  \label{eq:tau}
\end{equation}%
That is, by calculating the variance of the available first order effects
and halving we obtain an estimate of the total order indices for each input.
These estimates, compared with estimates of the first order indices provides
information on how much each input is involved in interaction with the
remaning ones. In particular, we note the
notion of mean effective dimension provides a quantitative indication about
the relevance of interactions. This notion has been introduced in \cite%
{CaflMoro97,Owen03}. The intuition is as follows. The terms $S_{i,j,...,r}=%
\dfrac{V_{i,j,...,r}}{V_{Y}}$ are positive and sum to unity. Thus, they can
be regarded as forming a probability mass function. Actually, they place a
mass on the random variable $T$ whose realizations all the possible
combinations of indices. Then, one defines the dimension distribution of $g$
in the superimposition sense as the distribution of the cardinality of $|T|$%
. The mean dimension is then \cite{Owen03}: 
\begin{equation}
D_{g}=\sum\nolimits_{|z|>0,z\subseteq \{1,2,...,n\}}|z|\Pr
(T=z)=\sum\nolimits_{|z|>0,z\subseteq \{1,2,...,n\}}|z|\dfrac{V_{z}}{V_{Y}}%
\text{,}  \label{e:D:T:Psi}
\end{equation}%
respectively. A mean effective dimension in the superimposition sense equal
to unity indicates the absence of interactions. \cite{Owen03} also proves
that the mean effective dimension is equal to the sum of total effects. That
is, we have: $D_{g}=\sum\nolimits_{i=1}^{n}T_{i}$. 

We recall that first order variance-based sensitivity measures fall in the
so-called common rationale of global sensitivity measures \cite{BorgHazePlish15}. That is, they are sensitivity measures of the type 
\begin{equation}
\xi _{i}=\mathbb{E}[d(P_{Y},P_{Y|X_{i}})],
\end{equation}%
where $d(\cdot ,\cdot )$ is a distance or divergence between distributions, $%
P_{Y}$ is the marginal probability measure of $Y$ and $P_{Y|X_{i}}$ is the
conditional probability measure of $Y$ given $X_{i}$. Depending on the
choice of $d(\cdot ,\cdot )$, one obtains alternative ways of measuring
importance. To illustrate, let $F_{Y}$ and $F_{Y|X_{i}}$ denote the
cumulative distribution function of $Y$ and the conditional distribution
function of $Y$ given $X_{i}$. Selecting the Kuiper distance between
cumulative distributions functions, we get \cite{Baucells2013} 
\begin{equation}
\beta _{i}^{Ku}=\mathbb{E}[\sup_{y}\{F_{Y|X_{i}}(y)-F_{Y}(y)\}+\sup_{y}%
\{F_{Y}(y)-F_{Y|X_{i}}(y)\}].  \label{eq:Ku}
\end{equation}%
We note that a null value of $T_{i}$ and $\beta _{i}^{Ku}$
indicates that $Y$ is independent of $X_{i}$. First order variance-based
sensitivity measure ($S_{i}$) do not possess this property and their null
value is not necessarily a signal of the fact that $ Y $ does not depend on $ X_i $. However, first order variance-based
sensitivity measures remain well defined when inputs are dependent,
similarly to $\beta _{i}^{Ku}$, while problems associated
with the interpretation of $T_{i}$ arise.

\subsection{Estimation}

The selected sensitivity measures for our calculations are the first order
main effects $\phi _{i}^{k\rightarrow k+1}$, the higher order effects $\phi
_{i,j,...,m}^{k\rightarrow k+1}$ computed at randomized locations. The
implementation of the decomposition in eqs. (\ref{eq:finitchangedec}) and (%
\ref{eq:phiijk}) is implemented in the surboutine \texttt{finitechanges.m}, which is
available at \url{https://github.com/LuXuefei/FiniteChanges}. Then, such decomposition is performed at $N$ randomized
locations appropriately generated through random Monte Carlo generation
implemented in the code \texttt{rand.m} and \texttt{randsample.m}, available at \textsc{Matlab}. 

From the randomized first order effects $\phi _{i}^{k\rightarrow k+1}$, it
is then possible to obtain estimates of the first order indices of all
inputs from 
\begin{equation}
\widehat{T}_{i}=\sum\limits_{k=1}^{N-1}\frac{(\phi _{i}^{k\rightarrow
		k+1})^{2}}{2}.
\end{equation}%

The estimation of the sensitivity measures $S_{i}$ and $\beta _{i}^{Ku}$ is
based on the so-called given data approach. We refer to \cite{PlisBorg13}
and \cite{BorgHazePlish15} for a detailed treatment. We just recall the key
intuition here. Let $\mathcal{X}_{i}$ a subset of the reals that represents
the support of model input $x_{i}$. Then, consider a partition of $\mathcal{X%
}_{i}$ into $M$ bins, such that $\cup _{m=1}^{M}\mathcal{X}_{i}^{m}=\mathcal{%
	X}_{i}$ and $\mathcal{X}_{i}^{m}\cap \mathcal{X}_{i}^{r}=\emptyset $ for $%
r,m=1,2,...,M$, $r\neq m.$ The intuition of a given-data estimation is then
that of substituting the condition $X_{i}=x_{i}$ (that is $X_{i}$ is exactly
at $x_{i}$) with the bin condition $X_{i}\in \mathcal{X}_{i}^{m}$.
Intuitively, this condition asks that $X_{i}$ is in a suitably defined
interval around $x_{i}$. With this intuition, one obtaines the estimate 
\begin{equation}
\widehat{\xi }_{i}=\sum_{m=1}^{M}p_{m,i}d(\widehat{F}_{Y},\widehat{F}%
_{Y|X_{i}\in \mathcal{X}_{i}^{m}}),
\end{equation}%
where $p_{m,i}$ is equal to the probability that $X_{i}$ belongs to bin $%
\mathcal{X}_{i}^{m}$, and where $\widehat{F}_{Y}$ and $\widehat{F}%
_{Y|X_{i}\in \mathcal{X}_{i}^{m}}$ are the marginal empirical distribution
function of $Y$ and the conditional empirical distribution function given
that $X_{i}\in \mathcal{X}_{i}^{m}$. Then, one links the partition size (M)
to the sample size ($N$), so that as $N$ increases the parition size tends
to zero and the limiting partition is $\mathcal{X}_{i}$ itself. \cite{BorgHazePlish15} prove that, under mild assumptions, (the continuity of the
operator $d(\cdot ,\cdot )$ in both its arguments and the Riemann-Stieltjes
integrability of $d(F_{Y},F_{Y|X_{i}})$), $\widehat{\xi }_{i}$ converges to $%
\xi _{i}$ as the sample size increases; that is, $\widehat{\xi }_{i}$ is an
asymptotically consistent estimator of $\xi _{i}$.

These calculations are implemented in the subroutine \texttt{betaKS3.m}. Finally, the conditional regression functions $r_{i}(x_{i})=\mathbb{E}%
[Y|X_{i}=x_{i}]$ have been estimated using the subroutine \texttt{cosi.m}. Both subroutines are developed
by Elmar Plischke and available at \href{https://zenodo.org/record/885332\#.XpnB44gzZGM}{https://zenodo.org/record/885332\#.XpnB44gzZGM}
(see \cite{Plischke2010} and \cite{Plis12EMS} for details).

\clearpage
\bibliographystyle{unsrt}


\begin{thebibliography}{10}
	
	\bibitem{Wu2020265}
	F~Wu, S~Zhao, B~Yu, Y.-M. Chen, W~Wang, Z.-G. Song, Y~Hu, Z.-W. Tao, J.-H.
	Tian, Y.-Y. Pei, M.-L. Yuan, Y.-L. Zhang, F.-H. Dai, Y~Liu, Q.-M. Wang, J.-J.
	Zheng, L~Xu, E~C Holmes, and Y.-Z. Zhang.
	\newblock {A new Coronavirus Associated with Human Respiratory Disease in
		China}.
	\newblock {\em Nature}, 579(7798):265--269, 2020.
	
	\bibitem{RamuRamu20}
	A.~Remuzzi and G.~Remuzzi.
	\newblock {COVID-19 and Italy: what next?}
	\newblock {\em The Lancet}, 395:1225--1228, 2020.
	
	\bibitem{Grasselli2020}
	G.~Grasselli, A.~Pesenti, and M.~Cecconi.
	\newblock {Critical Care Utilization for the COVID-19 Outbreak in Lombardy,
		Italy: Early experience and Forecast during An Emergency Response. JAMA.}
	\newblock {\em Journal of the American Medical Association}, 2020.
	
	\bibitem{Ande20Lancet}
	Roy~M. Anderson, Hans Heesterbeek, Don Klinkenberg, and T.~D{\'{e}}irdre
	Hollingsworth.
	\newblock {How will country-based mitigation measures influence the course of
		the COVID-19 epidemic?}
	\newblock {\em The Lancet}, 395(10228):931--934, 2020.
	
	\bibitem{Wu2020689}
	J~T Wu, K~Leung, and G~M Leung.
	\newblock {Nowcasting and forecasting the potential domestic and international
		spread of the 2019-nCoV outbreak originating in Wuhan, China: a modelling
		study}.
	\newblock {\em The Lancet}, 395(10225):689--697, 2020.
	
	\bibitem{EnseKupf20}
	M~Enserink and K~Kupferschmidt.
	\newblock {With COVID-19, modeling takes on life and death importance}.
	\newblock {\em Science}, 367(6485):1414--1415, 2020.
	
	\bibitem{Kucharski2020}
	A~J Kucharski, T~W Russell, C~Diamond, Y~Liu, J~Edmunds, S~Funk, R~M Eggo,
	F~Sun, M~Jit, J~D Munday, N~Davies, A~Gimma, K~van Zandvoort, H~Gibbs,
	J~Hellewell, C~I Jarvis, S~Clifford, B~J Quilty, N~I Bosse, S~Abbott,
	P~Klepac, S~Flasche, and Centre {for Mathematical Modelling of Infectious
		Diseases COVID-19 working group}.
	\newblock {Early dynamics of transmission and control of COVID-19: a
		mathematical modelling study}.
	\newblock {\em The Lancet Infectious Diseases}, 2020.
	
	\bibitem{Wang2020}
	H~Wang, Z~Wang, Y~Dong, R~Chang, C~Xu, X~Yu, S~Zhang, L~Tsamlag, M~Shang,
	J~Huang, Y~Wang, G~Xu, T~Shen, X~Zhang, and Y~Cai.
	\newblock {Phase-adjusted estimation of the number of Coronavirus Disease 2019
		cases in Wuhan, China}.
	\newblock {\em Cell Discovery}, 6(1), 2020.
	
	\bibitem{Peng2020}
	Liangrong Peng, Wuyue Yang, Dongyan Zhang, Changjing Zhuge, and Liu Hong.
	\newblock {Epidemic analysis of COVID-19 in China by dynamical modeling}.
	\newblock feb 2020.
	
	\bibitem{Salt19}
	A.~Saltelli.
	\newblock {A Short Comment on Statistical versus Mathematical Modelling}.
	\newblock {\em Nature Communications}, pages 1--2, 2019.
	
	\bibitem{Kermack1927}
	W.O. Kermack and A.G McKendrick.
	\newblock {A Contribution to the Mathematical Theory of Epidemics}.
	\newblock {\em Proceedings of the Royal Society A}, 115(772):700--721, 1927.
	
	\bibitem{Hethcote2000}
	Herbert~W. Hethcote.
	\newblock {Mathematics of infectious diseases}.
	\newblock {\em SIAM Review}, 2000.
	
	\bibitem{Cheynet2020}
	E~Cheynet.
	\newblock {Generalized SEIR Epidemic Model (fitting and computation)
		(https://www.github.com/ECheynet/SEIR)}, 2020.
	
	\bibitem{Bui2020}
	Quoctrung Bui, Josh Katz, Alicia Parlapiano, and Margot Sanger-Katz.
	\newblock {What 5 Coronavirus Models Say the Next Month Will Look Like}.
	\newblock {\em The New York Times}, apr 2020.
	
	\bibitem{Saltelli2008}
	A.~Saltelli, M.~Ratto, T.~Andres, F.~Campolongo, J.~Cariboni, D.~Gatelli,
	M.~Saisana, and S.~Tarantola.
	\newblock {\em {Global Sensitivity Analysis -- The Primer}}.
	\newblock Chichester, 2008.
	
	\bibitem{Sull15}
	T.J. Sullivan.
	\newblock {\em {Introduction to Uncertainty Quantification}}.
	\newblock Springer Verlag, 2015.
	
	\bibitem{Borgonovo2016book}
	E.~Borgonovo.
	\newblock {\em {Sensitivity Analysis: An Introduction for the Management
			Scientist}}.
	\newblock Springer, NY, 2017.
	
	\bibitem{Li2001}
	G.~Li, C.~Rosenthal, and H.~Rabitz.
	\newblock {High dimensional model representations}.
	\newblock {\em The Journal of Physical Chemistry A}, 105(33):7765--7777, 2001.
	
	\bibitem{Borgonovo2010}
	E.~Borgonovo.
	\newblock {Sensitivity analysis with finite changes: An application to modified
		EOQ models}.
	\newblock {\em European Journal of Operational Research}, 200(1):127--138,
	2010.
	
	\bibitem{EfroStei81}
	B.~Efron and C.~Stein.
	\newblock {The Jackknife Estimate of Variance}.
	\newblock {\em The Annals of Statistics}, 9(3):586--596, 1981.
	
	\bibitem{Homma1996}
	T.~Homma and A.~Saltelli.
	\newblock {Importance Measures in Global Sensitivity Analysis of Nonlinear
		Models}.
	\newblock {\em Reliability Engineering {\&} System Safety}, 52(1):1--17, 1996.
	
	\bibitem{Owen03}
	A.~B. Owen.
	\newblock {The Dimension Distribution and Quadrature Test Functions}.
	\newblock {\em Statistica Sinica}, 13:1--17, 2003.
	
	\bibitem{Sobo93}
	I.M. Sobol'.
	\newblock {Sensitivity Estimates for Nonlinear Mathematical Models}.
	\newblock {\em Mathematical Modelling {\&} Computational Experiments},
	1:407--414, 1993.
	
	\bibitem{GambJano16}
	F.~Gamboa, A.~Janon, T.~Klein, A.~Lagnoux, and C.~Prieur.
	\newblock {Statistical inference for Sobol pick-freeze Monte Carlo method}.
	\newblock {\em Statistics}, 50(4):881--902, 2016.
	
	\bibitem{BorgRabi20}
	E.~Borgonovo, G.~Rabitti
	\newblock {Simulator Inputs Screening: from Elementary Effects to
		Mean Dimensions}.
	\newblock {\em Work in Progress}, 1--39, 2020.
	
	\bibitem{CaflMoro97}
	R.~E. Caflisch, W.~Morokoff, and A.~B. Owen.
	\newblock {Valuation of mortgage backed securities using Brownian bridges to
		reduce effective dimension}.
	\newblock {\em Journal of Computational Finance}, 1:27--46, 1997.
	
	\bibitem{BorgHazePlish15}
	E.~Borgonovo, G.~Hazen, and E.~Plischke.
	\newblock {A Common Rationale for Global Sensitivity Measures and their
		Estimation}.
	\newblock {\em Risk Analysis}, 36(10):1871--1895, 2016.
	
	\bibitem{Baucells2013}
	M.~Baucells and E.~Borgonovo.
	\newblock {Invariant Probabilistic Sensitivity Analysis}.
	\newblock {\em Management Science}, 59(11):2536--2549, 2013.
	
	\bibitem{PlisBorg13}
	E.~Plischke, E.~Borgonovo, and C.L. Smith.
	\newblock {Global Sensitivity Measures from Given Data}.
	\newblock {\em European Journal of Operational Research}, 226(3):536--550,
	2013.
	
	\bibitem{Plischke2010}
	E~Plischke.
	\newblock {An effective algorithm for computing global sensitivity indices
		(EASI)}.
	\newblock {\em Reliability Engineering {\&} System Safety}, 95 (4):354--360,
	2010.
	
	\bibitem{Plis12EMS}
	E.~Plischke.
	\newblock {How to Compute Variance-Based Sensitivity Indicators with Your
		Spreadsheet Software}.
	\newblock {\em Environmental Modelling {\&} Software}, 35:188--191, 2012.
	
\end{thebibliography}

\end{document}